\documentclass[%
reprint, 
superscriptaddress,
amsmath,amssymb,
aps,
prl,
floatfix,
]{revtex4-1}

\usepackage[dvipsnames]{xcolor}

\usepackage{graphicx}
\usepackage{epsfig} 
\usepackage{dcolumn}
\usepackage{bm}
\usepackage{epstopdf}
\usepackage{multirow}
\usepackage{amsmath}
\usepackage{booktabs}
\usepackage{color}
\usepackage{gensymb}
\usepackage{kantlipsum}  
\usepackage{hyperref}
\usepackage{braket}
\usepackage{physics}
\usepackage{url}
\usepackage[utf8]{inputenc}
\usepackage[T1]{fontenc}
\usepackage{mathptmx}
\usepackage{lineno}
\usepackage{comment}
\usepackage{soul}
\usepackage[normalem]{ulem}
\hypersetup{
    hidelinks
}

\begin{document}

\title{Electrostriction in a Bose-Einstein Condensate of Dipolar Molecules}

\preprint{APS/123-QED}

\author{Haneul Kwak}
\thanks{These authors contributed equally to this work.}
\affiliation{Department of Physics, Columbia University, New York, New York 10027, USA}
\author{Ian Stevenson}
\thanks{These authors contributed equally to this work.}
\affiliation{Department of Physics, Columbia University, New York, New York 10027, USA}
\author{Weijun Yuan}
\affiliation{Department of Physics, Columbia University, New York, New York 10027, USA}
\author{Siwei Zhang}
\affiliation{Department of Physics, Columbia University, New York, New York 10027, USA}
\author{Asaf Toprak\c{c}\i}
\affiliation{Department of Physics, Columbia University, New York, New York 10027, USA}
\author{Lin Su}
\affiliation{Department of Physics, Columbia University, New York, New York 10027, USA}
\author{Tijs Karman}
\affiliation{Institute for Molecules and Materials, Radboud University, 6525 AJ Nijmegen, Netherlands}
\author{Sebastian Will}\email{Corresponding author. Email: sebastian.will@columbia.edu}
\affiliation{Department of Physics, Columbia University, New York, New York 10027, USA}
\date{\today}

\begin{abstract}
The recent creation of a Bose-Einstein condensate (BEC) of dipolar molecules has opened a new frontier for many-body quantum systems in which dipolar interactions can drive novel self-organization phenomena. Here, we observe \textit{electrostriction} in a molecular BEC, an elliptical deformation driven by anisotropic dipolar interactions. We use double microwave dressing, involving $\sigma$- and $\pi$-polarized fields, to control non-axially symmetric dipolar interactions. We compare the experimental observations of electrostriction to a model based on an extended Gross-Pitaevskii equation and find excellent agreement in the regime of weak to moderate interactions. Using electrostriction, we demonstrate that the molecular BEC can be torqued by dynamically changing the orientation of the elliptical $\sigma$ microwave field. This provides a route to setting molecular quantum gases into rotation, opening opportunities to probe vorticity, superfluidity, and supersolidity in strongly dipolar matter. 
\end{abstract}

\maketitle

\section{Introduction}

Quantum degenerate gases of dipolar molecules provide a novel platform for the exploration of many-body systems in which long-range interactions give rise to macroscopic quantum behavior~\cite{baranov2012condensed}. Based on the prospect of widely tunable dipolar interactions, theoretical proposals have envisioned
new many-body phases, from quantum crystals~\cite{buchler2007strongly} and exotic droplet phases~\cite{schmidt2022self} to Mott insulators with fractional filling~\cite{capogrosso2010quantum} and lattice spin liquids~\cite{yao2018quantum}.
Experimental progress has long been hindered by the presence of universal two-body loss in ultracold molecular gases~\cite{bause2023ultracold}. Recently, we have shown that double microwave dressing~\cite{karman2025double}, in which dipolar molecules are dressed by a circularly polarized $\sigma$ field and a linearly polarized $\pi$ field, can efficiently suppress inelastic losses while simultaneously enabling the flexible tuning of dipolar interactions, both in strength and anisotropy~\cite{yuan2025extreme}. This has enabled the creation of the first Bose-Einstein condensate (BEC) of dipolar molecules~\cite{bigagli2024observation, shi2026bose} and the first observation of molecular droplets in the regime of strong dipole-dipole interactions~\cite{zhang2026droplet}. These advances open novel experimental possibilities, while, at the same time, the development of theoretical models that take into account the characteristic interactions between microwave-dressed molecules is an exciting frontier~\cite{langen2025dipolar,jin2025bose,ciardi2025self,zhang2025quantum,zhang2025supersolid, schindewolf2025strongly,ciardi2026equilibrium,melero2026self}.

Over the past two decades, dipolar many-body systems have been extensively investigated in the context of weakly interacting magnetic atoms. Experimental breakthroughs include the observation of magnetostriction, the elliptical deformation of quantum gases of magnetic atoms~\cite{tang2018tuning, wenzel2018anisotropic, aikawa2014observation}; the formation of droplets~\cite{kadau2016observing, schmitt2016self, chomaz2016quantum}; and the realization of supersolidity~\cite{bottcher2019transient, chomaz2019long, tanzi2019observation}. On the theoretical side, modeling based on the extended Gross-Pitaevskii equation (eGPE)~\cite{lima2011quantum, petrov2015quantum} has been highly successful. The key addition in the eGPE over the standard Gross-Pitaevskii equation~\cite{gross1961structure, pitaevskii1961vortex} is the Lee-Huang-Yang (LHY) term~\cite{lee1957eigenvalues}, describing the leading beyond-mean-field correction arising from quantum fluctuations. However, whether the eGPE provides a suitable description of dipolar quantum gases of microwave-dressed molecules remains an open question~\cite{jin2025bose,zhang2025quantum,ciardi2025self}.

There are important qualitative differences between microwave-dressed dipolar molecules and magnetic atoms. In the eGPE framework, the short-range scattering physics is modeled as a contact interaction, which is mathematically described by a zero-range $\delta$-function potential whose strength is parametrized by the s-wave scattering length $a_{\textrm{s}}$. For magnetic atoms~\cite{chomaz2022dipolar}, this assumption is well justified by the clear separation between typical scattering and dipolar lengths, which are of order $100~a_0$, and interparticle distances, which are larger than $1{,}000~a_0$. For microwave-dressed molecules, however, the interaction potential features a repulsive hard-core radius and scattering lengths around $1{,}000~a_0$~\cite{bigagli2024observation}, and dipolar lengths that are tunable from zero to more than $10{,}000~a_0$~\cite{yuan2025extreme, zhang2026droplet}. These interaction length scales constitute a significant fraction of typical interparticle spacings of around $10{,}000~a_0$. Furthermore, microwave dressing gives access to dipolar interactions with distinct anisotropies, including axially symmetric conventional dipolar, antidipolar, and non-axially symmetric dipolar interactions~\cite{zhang2026droplet,yuan2025extreme,biswas2026controlled}. Determining whether the characteristic features of microwave-dressed molecules can be captured by the contact-interaction description of the eGPE~\cite{schindewolf2025strongly} requires a detailed comparison between experiment and theory.

In this work, we observe electrostriction in a BEC of microwave-dressed NaCs molecules, providing an ideal setting to benchmark the eGPE treatment. Electrostriction manifests itself as an elongation (contraction) along the direction of attractive (repulsive) dipolar interactions. We develop a theoretical model based on the extended Gross-Pitaevskii equation and compare its predictions with the experimentally observed cloud sizes and aspect ratios. We find excellent agreement in the range of weak to moderate dipolar interactions, up to dipolar lengths of $6{,}000~a_0$. Using electrostriction, we further demonstrate that dynamical control over the orientation of the dipole-dipole interactions can exert a torque on the molecular BEC and set it into rotation, opening a path to probing vorticity and superfluid properties.

\begin{figure}[!t]
    \centering
    \includegraphics[width=\linewidth]{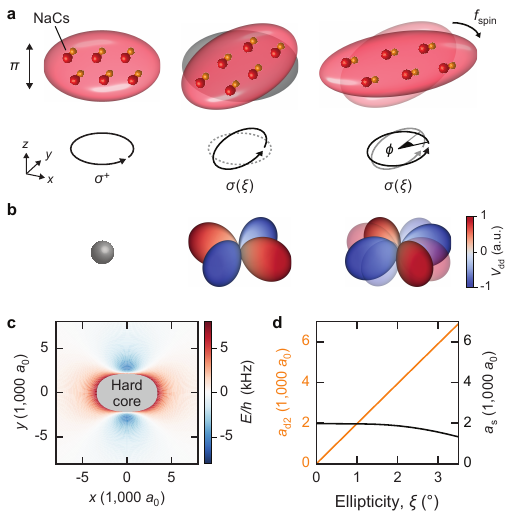}\\
    \caption{Electrostriction of a Bose-Einstein condensate of dipolar molecules. \textbf{a}, 
    Molecular interactions are controlled via double microwave dressing with $\sigma$ and $\pi$ dressing fields. For a circularly polarized $\sigma$ dressing field ($\xi=0^{\circ}$, left), dipolar interactions are compensated, and the shape of the BEC reflects the geometry of the trapping potential. For an elliptical $\sigma$ dressing field ($\xi>0^{\circ}$, center) the BEC shows an elongation and compression along the major and minor axes of the microwave ellipse, respectively. By rotating the polarization ellipse of the $\sigma$ field, electrostriction can be used to torque the BEC and set it into rotation (right). \textbf{b}, Angular dependence of the corresponding interaction potentials in \textbf{a}: hard-core potential ($\xi=0^\circ$, left), dipolar interactions ($\xi>0^\circ$, center), and dipolar interactions with rotating orientation (right). \textbf{c}, Molecular interaction potential in the $xy$ plane for $\xi=3^\circ$. Molecules experience attractive (repulsive) dipolar interactions along the major (minor) axis of the microwave ellipse. The gray shaded area denotes the hard-core potential truncated above 8 kHz. \textbf{d}, Dipolar length $a_{\textrm{d}2}$ and scattering length $a_{\textrm{s}}$ as a function of $\xi$, computed from the dressed state Hamiltonian and coupled-channels calculations, respectively (see Appendix).}
    \label{fig:1}
\end{figure}

\section{Microwave Dressing and Interactions}

Double microwave dressing allows the precise control of dipole-dipole interactions between molecules while providing a repulsive shield against inelastic loss~\cite{bigagli2024observation, karman2025double}. The molecules are simultaneously dressed with a circularly polarized $\sigma$ field and a linearly polarized $\pi$ field, as illustrated in Fig.~\ref{fig:1}\textbf{a}. The resulting dipole-dipole interaction potential can be expanded in spherical harmonics of multipole order $l=2$. The individual components $|m_l| = 0$, 1, and 2 have associated interaction length scales~\cite{karman2025double}, denoted by $a_{\mathrm{d0}}$, $a_{\mathrm{d1}}$, and $a_{\mathrm{d2}}$, respectively, which we individually control by independent microwave parameters (see Appendix). Specifically, $a_{\mathrm{d0}}$ is controlled by the amplitude balance of the $\sigma$ and $\pi$ fields~\cite{yuan2025extreme}, $a_{\mathrm{d1}}$ by the tilt between the axes of the $\sigma$ and $\pi$ fields~\cite{karman2025double}, and $a_{\mathrm{d2}}$ is controlled by $\xi$ denoting the ellipticity of the $\sigma$ field. In the experiments of this work, we choose $a_{\mathrm{d}0} \approx 0 $ and $a_{\mathrm{d}1} \approx 0$, which gives rise to non-axially symmetric dipole-dipole interactions as introduced in our earlier work~\cite{zhang2026droplet}. 

Via the $\sigma$-dressing field, we control both the strength and orientation of dipolar interactions. Experimentally, this is accomplished by controlling the ellipticity $\xi$ and the orientation of the polarization ellipse using the cloverleaf antenna \cite{yuan2023planar}. The $\sigma$ field is generated by combining two linearly polarized microwave fields, $E_{x'} \hat{x}'$ and $E_{y'} e^{i\phi} \hat{y}'$, with a controllable relative phase, $\phi$. Here, $\hat{x}'$ and $\hat{y}'$ denote the unit vectors of the microwave axes, which are slightly tilted with respect to the lab frame axes $x$ and $y$. Finite ellipticity can be generated by unbalancing $E_{x'}$ and $E_{y'}$ or by tuning away from $\phi=90^{\circ}$. Simultaneous control of phase and amplitude allows us to realize a polarization ellipse with arbitrary orientation (see Appendix). The microwave fields can be varied on fast time scales, enabling dynamic rotation of the polarization ellipse (see Fig.~\ref{fig:1}\textbf{a}).

Controlling the microwave ellipticity $\xi$ induces non-axially symmetric dipolar interactions~\cite{zhang2026droplet}, as illustrated in Figs.~\ref{fig:1}\textbf{b} and \ref{fig:1}\textbf{c}. The resulting long-range dipolar interactions can be expressed as
\begin{equation}
    V_{\rm dd}(\textbf{r}) = \dfrac{\sqrt{3}\hbar^2 }{mr^3} a_{\rm{d2}}\sin^2\theta \cos2\varphi.
    \label{eq:ddi}
    \nonumber
\end{equation}
Here, $m$ is the molecular mass and ${\bf r} = \{r, \theta, \varphi\}$ is the relative position vector between two molecules, where $\theta$ and $\varphi$ denote the polar and azimuthal angles of the intermolecular axis with respect to the $z$-axis, respectively. At short range, $r \lesssim 2{,}000~a_0$, microwave dressing induces an effective hard-core shield that is well described by a repulsive $1/r^6$ potential \cite{karman2025double,zhang2026droplet}. For the eGPE treatment, we approximate the short-range potential by an effective potential of the form
\begin{equation}
V_{\rm sr}(\textbf{r}) = g \delta(\textbf{r}),
\nonumber
\end{equation}
where $g = 4 \pi \hbar^2 a_\mathrm{s} / m$ parametrizes the strength of the contact interactions and $a_\mathrm{s}$ is the s-wave scattering length. Importantly, as shown in Fig.~\ref{fig:1}\textbf{d}, the s-wave scattering length is itself a function of the strength of the dipolar interactions~\cite{ronen2006dipolar}. In the absence of dipolar interactions for $\xi = 0$, $a_\mathrm{s}=2{,}000~a_0$. As dipolar interactions increase, $a_\mathrm{s}$ decreases monotonically. For $\xi > 5 \degree$, which is outside the range investigated in this work, $a_\mathrm{s}$ can even reach negative values.

We model the condensate dynamics using an eGPE that includes the effective potential
\begin{equation}
V_{\rm int}(\textbf{r}) = V_{\textrm{sr}}(\textbf{r}) +  V_{\rm dd}(\textbf{r})
\nonumber
\end{equation}
and a beyond-mean-field LHY correction term. For the non-axially symmetric dipolar interaction considered here, we derive the LHY correction from the anisotropic Bogoliubov spectrum of the system. When computing the LHY integral in momentum space, we introduce a direction-dependent lower momentum cutoff to exclude modes for which the Bogoliubov excitation energy becomes imaginary. This differs from the conventional approach of evaluating the LHY integral over the full momentum space and subsequently discarding its imaginary contribution. The resulting real-valued LHY term is used throughout our simulations. The eGPE model and the derivation of the LHY term are detailed in the Appendix.

\begin{figure}
    \centering
    \includegraphics[width=\linewidth]{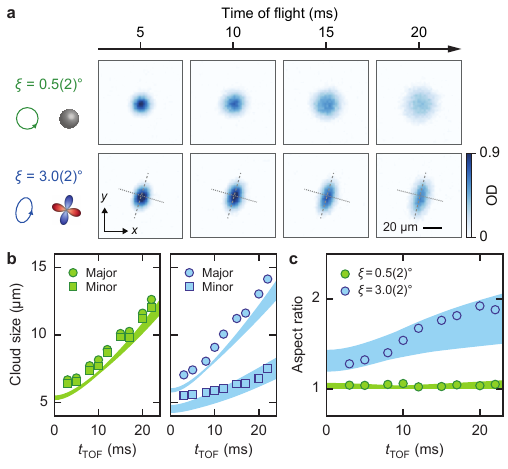}\\
    \caption{Observation of electrostriction in a molecular BEC. \textbf{a}, Time-of-flight expansion of a molecular cloud with $\xi = 0.5(2)^\circ$; $a_{\textrm{s}}=2{,}000~a_0$ and $a_{\textrm{d}2}=1{,}000~a_0$ (top row), and $\xi = 3.0(2)^\circ$; $a_{\textrm{s}}=1{,}600~a_0$ and $a_{\textrm{d}2}=5{,}900~a_0$ (bottom row). Pictures are the average of 15 to 20 experimental runs. The polarization ellipses (exaggerated for $\xi=3.0^\circ$) and the resulting angular dependence of the interaction potential are illustrated on the left. For $\xi=3.0^{\circ}$ the dashed and dotted gray lines highlight the cloud's major and minor axes, respectively. \textbf{b}, Cloud sizes ($1/\sqrt{e}$ radius) during time-of-flight expansion, obtained from two-dimensional Gaussian fits (see Appendix). Error bars are calculated as the statistical uncertainty from three repetitions of the experiment and are smaller than the data points. \textbf{c}, Aspect ratio of the clouds as a function of time of flight. The size of the BEC in this data set is $2.1(2) \times 10^3$ molecules. The shaded areas in panels \textbf{b} and \textbf{c} show the result of the eGPE simulation, accounting for the uncertainty in trap frequencies and the $a_{\mathrm{d0}}$ calibration.}
    \label{fig:2}
\end{figure}

\begin{figure*} [t]
    \centering
    \includegraphics[width=0.674157\linewidth]{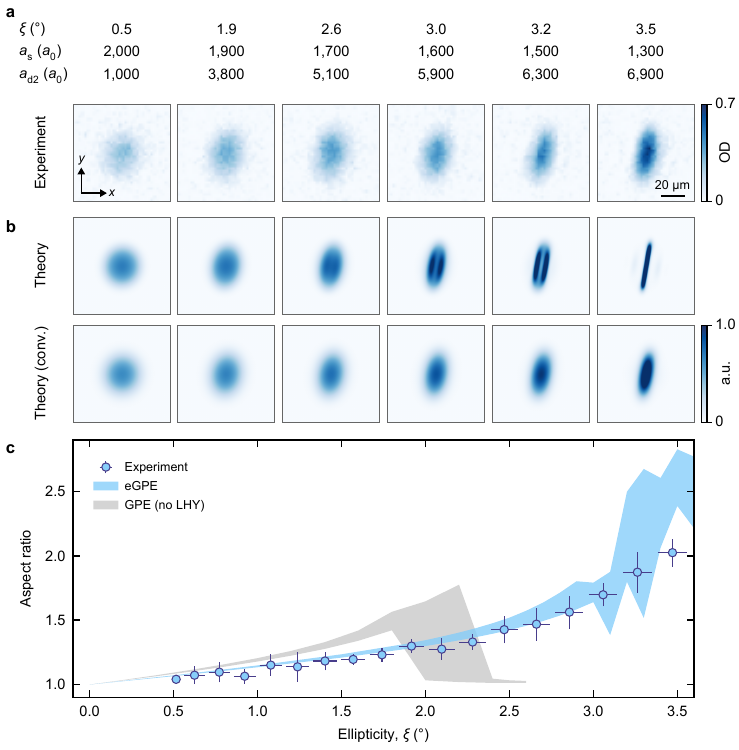}\\
    \caption{Electrostriction of a molecular BEC with tunable dipolar interactions. \textbf{a}, Density profiles of the BEC after 20 ms time-of-flight for various ellipticities. Each picture is the average of 5 experimental runs. \textbf{b}, Corresponding density profiles from eGPE simulation. Top row, raw images; bottom row, convolved images accounting for finite imaging resolution. The images are rotated clockwise by $9.6^\circ$ for comparison with the experiment. \textbf{c}, Aspect ratio of the cloud in the $xy$ plane as a function of ellipticity. The blue shaded area shows the result of the eGPE simulation. The gray shaded area shows the result of a standard GPE calculation without the LHY term. The size of the BEC in this data set is $1.7(3) \times 10^3$ molecules.}
    \label{fig:3} 
\end{figure*}

\begin{figure}
    \centering 
    \includegraphics[width=\linewidth]{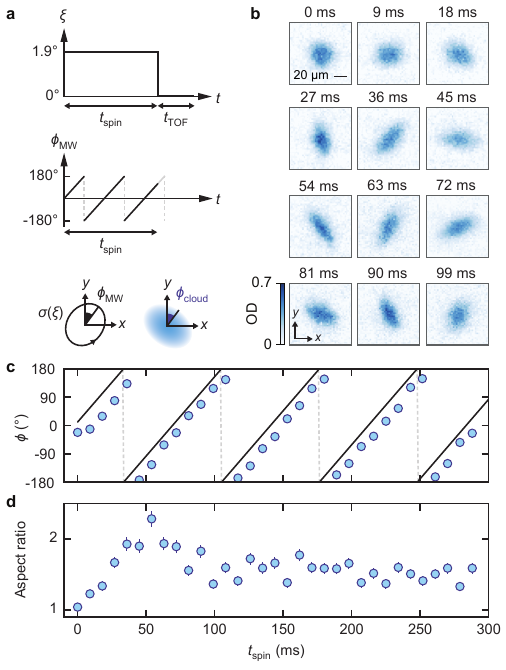}\\
    \caption{Using electrostriction to set a molecular BEC into rotation. \textbf{a}, Experimental sequence. The microwave ellipticity is kept at $\xi = 1.9^\circ$ while the orientation of the polarization ellipse is rotated at a frequency $f_{\textrm{spin}}$ for a variable spinning time $t_{\textrm{spin}}$, after which the ellipticity is quenched to zero and the non-dipolar cloud undergoes time-of-flight expansion during $t_{\textrm{TOF}}$. \textbf{b}, Time-of-flight images of the cloud for various spinning times. \textbf{c}, Orientation of the minor axis of the condensate as a function of spinning time (blue points). The black solid line shows the orientation of the microwave major axis. \textbf{d}, Condensate aspect ratio as a function of spinning time.}
    \label{fig:4} 
\end{figure}

\section{Electrostriction}

The experiment starts with a BEC of $2 \times 10^3$ NaCs molecules with a condensate fraction greater than 70\%. The molecules are held in an oblate trap that is round in the $xy$ plane, with trap frequencies $\omega_{x,y,z}/(2 \pi)  = \{21(1), 21(1), 48(4)\}$~Hz. The peak density of the BEC is about $3\times10^{12}\,\rm{cm}^{-3}$. The BEC is created in the regime of minimal $\sigma$-field ellipticity, where the dipolar interactions are minimized \cite{bigagli2024observation}. Experimentally, we reach $\xi = 0.5(2)^\circ$, corresponding to the interaction parameters $\{ a_\mathrm{s}, a_{\mathrm{d0}}, a_{\mathrm{d1}}, a_{\mathrm{d2}} \} = \{2{,}000_{-200}^{+100}, 0(900), 0(600), 900(360) \}~a_0$. Details on sample preparation and detection are described in earlier work~\cite{stevenson2023ultracold, warner2023efficient, bigagli2024observation}. At the end of evaporative cooling, the ellipticity $\xi$ is linearly ramped to a target value between $0.5^\circ$ and $3.5^\circ$, at a rate of 0.03$^\circ$/ms~\cite{zhang2026droplet}. After the ramp, the cloud is released from the trap and, following time-of-flight expansion, the density profile is recorded using absorption imaging along the $z$ axis. As the trap is round in the $xy$ plane, any anisotropy that is observed in the cloud profile can be attributed to dipolar interactions.

First, we study the evolution of the cloud shape during time-of-flight expansion, comparing the non-dipolar case to the case of finite dipolar interactions. Example images of expanding molecular clouds are shown in Fig.~\ref{fig:2}\textbf{a}. For minimal dipolar interactions, $\xi = 0.5^\circ$, the cloud expands isotropically. For $\xi = 3.0^\circ$, dipolar electrostriction leads to an elliptical cloud whose anisotropy increases further during time-of-flight expansion. The tilt of the elliptical cloud reveals the orientation of the microwave ellipse with respect to the lab frame. For quantitative analysis, we fit the clouds to rotated two-dimensional Gaussians, from which we obtain the cloud sizes, as plotted in Fig.~\ref{fig:2}\textbf{b}. The data show that the elliptical expansion is slightly faster along its major axis and significantly slower along the minor axis compared to the non-dipolar isotropic cloud. 

The evolution of the aspect ratio for $\xi = 3.0^\circ$ further highlights the distinct expansion dynamics of the dipolar BEC, see Fig.~\ref{fig:2}\textbf{c}. A non-dipolar BEC with contact interactions, prepared in an anisotropic trap, shows a characteristic inversion of its aspect ratio after time-of-flight expansion~\cite{castin1996bose}. In contrast, the dipolar BEC already shows an \textit{in situ} aspect ratio larger than one due to dipole-dipole interactions, even when prepared in a trap that is isotropic in the $xy$ plane. During time-of-flight expansion, the aspect ratio increases further as the anisotropic dipolar interactions enhance the deformation.

The agreement between the experimental and theoretical data for cloud size and aspect ratio is excellent. For the absolute cloud size, see Fig.~\ref{fig:2}\textbf{b}, experiment and theory exhibit small quantitative differences; the expansion rate of the cloud is about $10\%$ faster in the experiment than predicted by the eGPE. This discrepancy may arise from higher-order quantum fluctuations~\cite{tan2008three} or interactions between the BEC and the residual thermal cloud at finite temperature. For the cloud aspect ratio, the agreement is excellent as shown in Fig.~\ref{fig:2}\textbf{c}. The $\pm 900~a_0$ uncertainty in residual $a_{\rm{d}0}$ interactions is the dominant source of theoretical uncertainty, corresponding to about $\pm 15\%$ in the predicted aspect ratio.

Next, we study the cloud shape as a function of ellipticity $\xi$ at a fixed time-of-flight, as shown in Fig.~\ref{fig:3}. Both the experimental images (Fig.~\ref{fig:3}\textbf{a}) and the corresponding eGPE simulations (Fig.~\ref{fig:3}\textbf{b}) show that electrostriction becomes increasingly pronounced with increasing $\xi$. For $\xi = 3.5^\circ$, the aspect ratio reaches 22.5 in the simulation, while the experimentally measured aspect ratio is 2.0(2). After taking into account the finite imaging resolution (see Appendix), the experimental and simulated cloud shapes, as well as the resulting aspect ratios (Fig.~\ref{fig:3}\textbf{c}), show excellent agreement. Remarkably, the simulated density profiles show the onset of a density-modulated phase for $3^\circ\le\xi\le3.5^\circ$, forming two to three droplets that are connected by a bridge with finite density. The emergence of these density-modulated structures leads to large fluctuations in the simulated aspect ratio. In the experiment, however, they are not detected due to finite imaging resolution.

The observations in Fig.~\ref{fig:3} suggest that, in the BEC regime, the Gross-Pitaevskii equation with the LHY term as the leading beyond-mean-field correction provides an accurate theoretical description of electrostriction. To highlight the effect of quantum fluctuations, we show the result of a standard GPE simulation without the LHY term as the gray shaded area in Fig.~\ref{fig:3}\textbf{c}. Above $\xi = 2^\circ$, the standard GPE predicts a mean-field collapse, indicated by the rapid shrinking of the cloud below the imaging resolution and the aspect ratio approaching unity. Above $\xi = 2.6^\circ$, the standard GPE no longer converges. In contrast, the eGPE quantitatively agrees with the experimental observations across the interaction range explored in this work, from the nearly non-dipolar regime at $\xi = 0.5^\circ$ with $na_{\textrm{s}}^3 = 0.004$ and $na_{\textrm{d}2}^3 = 0.0004$ to the most strongly dipolar regime at $\xi = 3.5^\circ$ with $na_{\textrm{s}}^3 = 0.001$ and $na_{\textrm{d}2}^3 = 0.146$, for a typical peak condensate density $n = 3 \times 10^{12}\,\textrm{cm}^{-3}$.

Comparing our treatment of the LHY term with the conventional approach, in which the imaginary contribution is discarded, we find only marginal differences, even though the ratio of the imaginary to the real part of the conventional LHY term reaches approximately 7\% at $\xi=3.5^{\circ}$. However, beyond $\xi=3.5^{\circ}$, the imaginary contribution increases rapidly, and the precise value of the LHY term becomes increasingly sensitive to the details of the calculation (see Appendix). 

\section{Rotating Molecular BECs}

We demonstrate that electrostriction can be used to set a molecular BEC into rotation (see Fig.~\ref{fig:4}). As described above, the orientation of the microwave polarization ellipse determines the orientation of the dipolar interaction potential. By dynamically rotating the polarization ellipse and, with it, the dipolar interaction pattern, we can torque the electrostricted molecular BEC. For the data in Fig.~\ref{fig:4}\textbf{b}, we rotate the microwave ellipse clockwise at a frequency $f_{\textrm{spin}}=14$ Hz while the ellipticity stays constant at $\xi = 1.9^\circ$. After a spinning time $t_{\textrm{spin}}$, the ellipticity is quenched to zero, the cloud is released from the trap, and subsequently expands for 20~ms under isotropic interactions before absorption imaging.

We analyze the dynamics from its onset until the cloud reaches a steady-state rotation (Figs.~\ref{fig:4}\textbf{c} and \ref{fig:4}\textbf{d}). As the cloud spins up, its aspect ratio initially increases from near unity to $2.3$, as a result of centrifugal deformation, and then relaxes to a steady-state value near $1.5$ for longer $t_{\mathrm{spin}}$, as shown in Figs.~\ref{fig:4}\textbf{b} and \ref{fig:4}\textbf{d}. After $t_{\rm{spin}}=100$ ms, the elliptical cloud follows the rotating microwave ellipse with a phase lag of about $47(2)^{\circ}$. The relaxation of the aspect ratio after the initial overshoot is qualitatively consistent with vortex nucleation dynamics in rotating superfluids~\cite{madison2001stationary,Gallemi2020vortices, klaus2022observation}. In the current experimental setup, however, we cannot detect the potential appearance of vortices. \textit{In situ}, the diameter of vortex cores is on the scale of the healing length, $1/\sqrt{8 \pi n a_{\textrm{s}}} \approx 0.35~$\textmu m. During time-of-flight, the vortex cores should expand by the same factor as the overall BEC size~\cite{abo2001observation}, yielding an estimated size of $5 \times 0.35$ \textmu m $\sim 1.75~$\textmu m, which is below the imaging resolution of our current setup. The observation of vortex formation in a molecular BEC is an important goal for future experiments with improved imaging resolution.

\section{Conclusions and Outlook}

We have observed electrostriction in a molecular BEC, a deformation of the condensate driven by anisotropic dipolar interactions. We find excellent agreement between the experimentally measured cloud shape and the eGPE predictions for dipolar lengths ranging from near zero to $a_{\rm{d}2}=6{,}900~a_0$. This agreement indicates that the $\delta$-function description of interactions within the eGPE formalism quantitatively describes electrostriction in a molecular BEC. It also suggests that quantum fluctuations play a crucial role in the stabilization of the system. Finally, we observe the rotation of a molecular BEC driven by electrostriction as the microwave polarization ellipse is dynamically rotated.

Beyond the interaction range investigated in this work, the system undergoes droplet formation for $\xi > 3.5^\circ$~\cite{zhang2026droplet}. In this regime, the applicability of the eGPE remains an open question as the imaginary component of the conventional LHY term grows beyond 10\% of its real component (see Appendix). For $\xi>5^\circ$, the s-wave scattering length assumes negative values and the eGPE cannot describe the observation of stable molecular gases in this regime. However, the intermediate regime, $3.5^\circ < \xi < 5 ^\circ$, can provide an ideal testbed for further benchmarking of theoretical approaches, including the eGPE studied here, variational methods~\cite{jin2025bose}, and quantum Monte Carlo simulations~\cite{langen2025dipolar,zhang2025supersolid,ciardi2025self}.

Experimentally, we expect electrostirring to become a valuable tool for molecular BECs, similar to magnetostirring in quantum gases of magnetic atoms~\cite{ferrier2018scissors, klaus2022observation, casotti2024observation, poli2025synchronization}, but with extended experimental capabilities. Electrostirring is naturally compatible with existing microwave-dressing hardware, can be dynamically controlled with high precision and speed, and allows simultaneous tuning of dipole-dipole interaction strength and anisotropy. Controlled rotational excitation and vortex nucleation in molecular quantum liquids open the door to studies of molecular superfluidity, supersolids, and self-bound droplets under rotation.

\section{Acknowledgments}

We acknowledge Jason Ho, Tilman Pfau, Luis Santos, Reuben Wang, and Wilhelm Zwerger for helpful discussions. We are grateful to Aden Lam, Claire Warner, and Niccol\'o Bigagli for important contributions in the construction of the experimental apparatus. This work was supported by an NSF CAREER Award (Award No.~1848466), an NSF Single Investigator Award (Award No.~2409747), an ONR DURIP Award (Award No.~N00014-21-1-2721), an AFOSR Single Investigator Award (Award No.~FA9550-25-1-0048), and a grant from the Gordon and Betty Moore Foundation (Award No.~GBMF12340). T.K.~acknowledges NWO VIDI (Grant ID 10.61686/AKJWK33335). I.S.~was supported by the Ernest Kempton Adams Fund. S.W.~acknowledges additional support from the Alfred P. Sloan Foundation. 


%

\newpage
\renewcommand{\thefigure}{A\arabic{figure}}

\setcounter{figure}{0}

\section{Appendix}

\subsection{Derivation of dipolar interaction potentials}

To obtain the long-range dipolar interaction potentials between microwave-dressed dipolar molecules, we numerically diagonalize the dressed-state Hamiltonian
\begin{gather*} \label{eq:mat}
    H/\hbar = \begin{bmatrix}
        0 & \Omega_\sigma / 2 \cos(\xi) & \Omega_\pi / 2 & \Omega_\sigma / 2 \sin(\xi)\\
        \Omega_\sigma / 2 \cos(\xi) & -\Delta_\sigma & 0 & 0 \\
        \Omega_\pi / 2 & 0 & -\Delta_\pi & 0 \\
        \Omega_\sigma / 2 \sin(\xi) & 0 & 0 & -\Delta_\sigma
    \end{bmatrix}.
\end{gather*}
Diagonalization yields three bright eigenstates. The highest energy state $\ket{s}$ is the shielded state, while the two additional eigenstates are anti-shielded and not protected from two-body loss. For the $\ket{s} = \alpha \ket{0, 0} + \beta \ket{1, 1} + \gamma \ket{1,0} + \delta \ket{1,-1}$ eigenvector, the induced dipolar interaction potential is~\cite{karman2025double}
\begin{equation}
    V_{\mathrm{dd}} = \dfrac{2\hbar^2}{mr^3} \left(a_{\mathrm{d}2} \dfrac{C_{2,-2} + C_{2,2}}{\sqrt{2}} - a_{\mathrm{d}0} C_{20}\right)
    \nonumber
\end{equation}
with corresponding dipolar lengths
\begin{gather}
    a_{\mathrm{d}0} = \dfrac{m d_{0}^2}{4\pi\epsilon_0 \hbar^2} \dfrac{|\alpha|^2(2|\gamma|^2-|\beta|^2 - |\delta|^2)}{3},\nonumber\\
    a_{\mathrm{d}2} = \dfrac{m d_{0}^2}{4\pi\epsilon_0 \hbar^2} \dfrac{2 |\alpha|^2 \beta \delta}{\sqrt{3}}.\nonumber
\end{gather}
Here, $\alpha$, $\beta$, $\gamma$, and $\delta$ denote complex amplitudes which are controlled by the Rabi frequencies and detunings of the dressing fields, and $d_0$ is the permanent dipole moment of the NaCs molecule, 4.6~Debye~\cite{dagdigian1972molecular}. For ellipticities below $10^\circ$, $a_{\textrm{d}2} \propto \sin{2\xi}$ is an excellent approximation~\cite{zhang2026droplet}.

\subsection{Cloud fitting}

To extract the cloud sizes, we fit the absorption images with a rotated two-dimensional Gaussian of the form
\begin{align*}
n(x,y) = \frac{N}{2 \pi \sigma_x \sigma_y} e^{-a(x-x_0)^2} e^{-2b (x-x_0)(y-y_0)} e^{-c (y-y_0)^2},
\end{align*}
where $a = \cos^2 \theta / 2\sigma_x^2 + \sin^2 \theta/2 \sigma_y^2$, $b = \sin2\theta / 4 \sigma_x^2 - \sin 2\theta/4 \sigma_y^2$, $c = \sin^2 \theta/2\sigma_x^2 + \cos^2\theta/2 \sigma_y^2$. Here, $\sigma_x$ and $\sigma_y$ denote the minor- and major-axis cloud sizes, respectively, and $\theta$ is the tilt with respect to the $x$ axis. 

When the eGPE is used to generate images for comparison with the experiment, we first calculate the three-dimensional profile of the cloud and integrate along the $z$ axis to obtain a two-dimensional density image. We then convolve the image with a two-dimensional Gaussian with $1/\sqrt{e}$ radius $3.5$ \textmu m, corresponding to the imaging resolution of our system~\cite{zhang2026droplet}. From the simulated images we obtain cloud sizes and aspect ratios using the same fitting routine as for the experimental data.

\subsection{Extended GPE}

For theoretical modeling, we use the extended Gross-Pitaevskii equation (eGPE)~\cite{lima2012beyond, santos2016dropletgs},
\begin{equation*}
    i\hbar \dfrac{\partial\psi}{\partial t} = \left(H_0 + \int d^3 r' V_{\rm int}({\bf r} - {\bf r}') |\psi (\textbf{r}', t)|^2 + g_\text{LHY} |\psi|^3 \right) \psi. \label{eGPE}
\end{equation*}
Here, $\psi$ is the condensate wavefunction, normalized to the molecule number $N=\int d^3 r |\psi|^2$, and $H_0 = -\frac{\hbar^2}{2m}\nabla^2 + V_{\text{ext}}(\textbf{r})$ is the single-particle Hamiltonian with external harmonic trapping potential $V_{\text{ext}}(\textbf{r})=\frac{1}{2}m(\omega_x^2 x^2 + \omega_y^2 y^2 + \omega_z^2 z^2)$. The term $V_{\rm int}$ denotes the effective interaction potential between the microwave-dressed molecules, as discussed below. The term $g_\text{LHY} |\psi|^3$ represents the beyond-mean-field Lee-Huang-Yang (LHY) correction to the chemical potential arising from quantum fluctuations~\cite{lee1957eigenvalues}. 

For the particular form of the non-axially symmetric dipolar interactions used in this work~\cite{zhang2026droplet}, we derive the coupling constant $g_\text{LHY}$. Following the treatment of Ref.~\cite{lima2012beyond}, we obtain
\begin{equation*}
    g_\text{LHY} = \frac{32}{3}g\sqrt{\frac{a_{\textrm{s}}^3}{\pi}}Q_5
    \label{LHY}
\end{equation*}
where
\begin{equation*}
    \label{eq:q5}
    Q_5 = \frac{1}{4\pi} \int_0^\pi d\theta \int_0^{2\pi} d\phi \sin{\theta} \left(1 - \epsilon_2 \sin^2\theta \cos 2\phi\right)^{5/2}.
\end{equation*}
When the term in parentheses takes negative values for certain angles, $Q_5$ acquires an imaginary component. The conventional approach is to neglect this imaginary component \cite{chomaz2022dipolar, ferrier2016observation}. In our calculations, however, we use the modified real-valued LHY term derived in the section \textbf{Lee-Huang-Yang (LHY) term}. Over the ellipticity range considered in this work, the imaginary part of the conventional $Q_5$ term is at most about $7\%$ of its real part. Here we define $\epsilon_{2} =\sqrt{8/3} (a_{\textrm{d}2} / a_{\textrm{s}})$ for convenience. 

\subsection{Lee-Huang-Yang (LHY) term}

To derive the LHY correction to the chemical potential for the non-axially symmetric dipolar interaction considered here, we follow the treatment of Ref.~\cite{lima2012beyond}. Within the local density approximation, the renormalized correction to the ground-state energy $\Delta E$ is
\begin{equation}
    \Delta E = \frac{1}{2}V\int\frac{\text{d}^3 k}{(2\pi)^3}\left[ E_\textbf{k} - \frac{\hbar^2\textbf{k}^2}{2m} - n_0 \tilde{V}_\textrm{int}(\textbf{k}) + \frac{\tilde{V}^2_\textrm{int}(\textbf{k})}{2k^2}\frac{2mn_0^2}{\hbar^2}\right], \label{E_correction}
\end{equation}
where $V$ is the quantization volume of the locally homogeneous system and
\begin{equation}
    E_\textbf{k} = \sqrt{ \epsilon_{\textbf{k}}\left( \epsilon_{\textbf{k}} + 2n_0\tilde{V}_\textrm{int}(\textbf{k}) \right)},\label{E_q}
\end{equation}
is the Bogoliubov excitation energy. Here, 
$\epsilon_{\textbf{k}} = \frac{\hbar^2 \textbf{k}^2}{2m}$ is the free particle energy, $n_0$ the condensate density, and $\tilde{V}_\textrm{int}(\textbf{k})=\int d^3 r e^{i \textbf{k}\cdot\textbf{r}}V_{\textrm{int}}(\textbf{r})$ the Fourier transform of the total interaction potential, with
\begin{equation*}
    \tilde{V}_\textrm{int}(\textbf{k}) = g + \tilde{V}_{\textrm{dd}}(\textbf{k}).
\end{equation*}
Note that $\tilde{V}_\textrm{int}(\textbf{k})$ only depends on the direction of $\textbf{k}$ which is characterized by the angles $\theta_{\textbf{k}}$ and $\phi_{\textbf{k}}$. Evaluating the local energy density correction gives
\begin{equation}
    \Delta E = V g n_0 ^2 \frac{64}{15}\sqrt{\frac{n_0 a_{\textrm{s}}^3}{\pi}}Q_5 \label{E_ground}
\end{equation}
where
\begin{equation*}
    Q_5 = \dfrac{1}{4\pi}\int d\Omega_{\hat{\textbf{k}}}\left[\tilde{V}_{\textrm{int}}(\textbf{k}) / g\right]^{5/2}
\end{equation*}
and hence 
\begin{equation*}
    \mu_\text{QF} = \frac{32gn_0 }{3}\sqrt{\frac{n_0 a_{\textrm{s}}^3}{\pi}}Q_5.
\end{equation*}

From Eqs. \eqref{E_correction} and \eqref{E_q}, we find that $E_{\textbf{k}}$ becomes imaginary for a range of momenta $\textbf{k}$. In earlier studies, the imaginary contribution to the LHY integral was simply discarded for relatively small interaction strengths, but the validity of this treatment is unclear at stronger interactions, particularly in the antidipolar regime. More recent work uses a momentum cutoff in Eq.~\eqref{E_correction} such that the integrand excludes unstable modes $\textbf{k}$ with imaginary values of $E_{\textbf{k}}$~\cite{oktel2019temperature}. We therefore modify the LHY term by retaining only stable modes, as described below.

Introducing a direction-dependent energy cutoff, corresponding to a momentum cutoff for each direction $\hat{\textbf{k}}$, the ground state energy correction in \eqref{E_correction} becomes
\begin{equation*}
\begin{aligned}
\Delta E
&= \frac{\sqrt{2}V}{4\pi^2}\frac{m^{3/2}}{\hbar^3}
   \int \frac{d\Omega_{\hat{\mathbf{k}}}}{4\pi}
   \int_{\epsilon_0(\hat{\mathbf{k}})}^{\infty} d\epsilon\, \sqrt{\epsilon}\,
   \Bigl[\sqrt{\epsilon^2 + 2n_0 \tilde{V}_{\mathrm{int}}(\mathbf{k})\,\epsilon}
\\
&\quad\quad\quad\quad\quad\quad\quad\quad\quad
   - \epsilon - n_0 \tilde{V}_{\mathrm{int}}(\mathbf{k})
   + \frac{(n_0 \tilde{V}_{\mathrm{int}}(\mathbf{k}))^2}{2\epsilon} \Bigr]
\\
&= \frac{\sqrt{2}V}{4\pi^2}\frac{m^{3/2}}{\hbar^3}
   \int \frac{d\Omega_{\hat{\mathbf{k}}}}{4\pi}\, I(\hat{\mathbf{k}}),
\end{aligned}
\end{equation*}
where
\begin{equation*}
\begin{aligned}
    I(\hat{\textbf{k}}) &\equiv \dfrac{1}{15\sqrt{\epsilon_0}}\Bigl[-15(n_0 \tilde{V}_{\textrm{int}}(\textbf{k}))^2\epsilon_0 + 6\epsilon_0^3+10 n_0 \tilde{V}_{\textrm{int}}(\textbf{k}) \epsilon_0^2
    \\
    &\quad\quad\quad\quad\quad+ (16(n_0 \tilde{V}_{\textrm{int}}(\textbf{k}))^2 - 6 \epsilon_0^2 - 4n_0 \tilde{V}_{\textrm{int}}(\textbf{k}) \epsilon_0)
    \\
    &\quad\quad\quad\quad\quad\quad\quad\quad\times\sqrt{\epsilon_0 (2 n_0 \tilde{V}_{\textrm{int}}(\textbf{k}) + \epsilon_0)}\Bigr].
\end{aligned}
\end{equation*}
Here, $\epsilon_0(\hat{\textbf{k}})$ is the direction-dependent energy cutoff. In this work, we choose $\epsilon_0 (\hat{\textbf{k}}) = \text{max}\{0, -2n_0 \tilde{V}_{\textrm{int}}(\textbf{k})\}$, which excludes the complex integrand at unstable modes. Then the ground state energy correction \eqref{E_ground} can be expressed by a modified $Q'_5$ function
\begin{equation*}
    Q'_5 = \dfrac{1}{4\pi}\int d\Omega_{\hat{\textbf{k}}} f_5 (\hat{\textbf{k}}),
\end{equation*}
with the integrand $f_5(\hat{\textbf{k}})$ being
\begin{equation*}
    f_5 (\hat{\textbf{k}}) = 
    \begin{cases}
    \left[\tilde{V}_{\textrm{int}}(\textbf{k})/{g}\right]^{5/2} & \text{if  } \tilde{V}_{\textrm{int}}(\textbf{k}) \ge 0  \\
    -\dfrac{11}{16}\left[-\tilde{V}_{\textrm{int}}(\textbf{k})/{g} \right]^{5/2}& \text{if  } \tilde{V}_{\textrm{int}}(\textbf{k})<0.
    \end{cases}
\end{equation*}
The prime distinguishes this modified real-valued function from the conventional $Q_5$. 

\begin{figure}
    \centering
    \includegraphics[width=\linewidth]{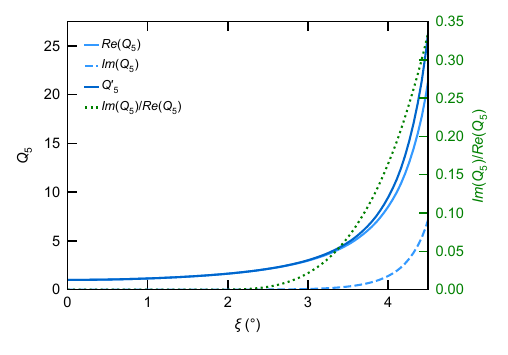}
    \caption{$Q_5$ coefficients of the LHY term. The real and imaginary parts of the conventional $Q_5$ term are plotted as the solid and dashed blue lines, respectively. The dark blue line shows the modified real-valued $Q'_5$ term. The ratio $\mathrm{Im}(Q_5)/\mathrm{Re}(Q_5)$ is plotted on the right axis (green dotted line).}
    \label{fig:FIGS4}
\end{figure}

Figure~\ref{fig:FIGS4} compares the real and imaginary parts of the conventional $Q_5$ term with our modified real-valued $Q'_5$ term. Within the ellipticity range explored in this work, these two approaches differ only marginally. For $\xi>3.5^\circ$, however, the imaginary part of the conventional $Q_5$ becomes increasingly more significant compared to its real part.

\subsection{Orienting the microwave ellipse}

By tuning $E_{x'}$, $E_{y'}$, and $\phi$, we control both the 
ellipticity and orientation of the elliptical $\sigma$ field. When $\phi = 90^\circ$ and $E_{x'} = E_{y'}$, the field is perfectly circular, corresponding to $\xi = 0$. For the special case $\phi = 90^\circ$ and $E_{x'}\neq E_{y'}$, the ellipticity is $\xi=\arctan[(E_{x'} - E_{y'}) / (E_{x'} + E_{y'})]$. For the general case $\phi \neq 90^\circ$, we define $\tau = \arctan (E_{x'} / E_{y'})$. The ellipticity is then determined by solving $\cos{2\xi} = \sin{\phi} \sin{2 \tau}$, and the tilt angle $\theta$ of the polarization ellipse can be found by solving $(\tan{2 \theta} / \tan{2 \tau})^2 + (\cos{2 \xi} / \sin{2 \tau})^2 = 1$.

\end{document}